# Security Incident Response Criteria: A Practitioner's Perspective

*Full paper*

**George Grispos**
University of Glasgow
g.grispos.1@research.gla.ac.uk

**William Bradley Glisson**
University of South Alabama
bglisson@southalabama.edu

**Tim Storer**
University of Glasgow
Timothy.Storer@glasgow.ac.uk

## Abstract[1]

Industrial reports indicate that security incidents continue to inflict large financial losses on organizations. Researchers and industrial analysts contend that there are fundamental problems with existing security incident response process solutions. This paper presents the Security Incident Response Criteria (SIRC) which can be applied to a variety of security incident response approaches. The criteria are derived from empirical data based on in-depth interviews conducted within a Global Fortune 500 organization and supporting literature. The research contribution of this paper is twofold. First, the criteria presented in this paper can be used to evaluate existing security incident response solutions and second, as a guide, to support future security incident response improvement initiatives.

## *Keywords*

Security Incident Response, Cybercrime, Security Incident Response Criteria (SIRC).

## Introduction

Industry reports indicate that security incidents are continuing to inflict staggering financial losses on organizations. The 2014 Ponemon Cost of Cyber Crime (2014b) report estimates that cybercrime costs organizations in the United States an average of $12.7 million, an increase of 9% since the previous year. In concert with these findings, the 2014 PricewaterhouseCoopers Global State of Information Security Survey (2014) reports that the number of respondents who reported losses of more than $10 million from security incidents increased by 51% since 2011. In addition to financial losses from security incidents, additional regulatory pressure has been applied to organizations in a variety of industries, mandating information security incident handling requirements (Grama 2014). For example, within the healthcare industry, the introduction of the 'Security Rule' to the Health Insurance Portability and Accountability Act of 1996 dictates organizations implement policies and procedures to detect, contain, and correct security violations (Johnson 2013b). Similarly, the Federal Information Security Management Act of 2002, and Sarbanes-Oxley Act of 2002 both require organizations to have policies and procedures to detect, report and respond to security incidents (Johnson 2013b).

The European Network and Information Security Agency (2010) and the National Institute of Standards and Technology (Cichonski et al. 2012) have published documentation guiding security incident response teams during the detection, investigation and eradication of security incidents. In addition, various security incident response approaches have been proposed by academia (Mandia et al. 2003; Mitropoulos et al. 2006). However, recent discussions have argued that significant problems exist with the application of these approaches not coping with time pressures, incident prioritization, in-depth root cause analysis and overall incident learning (FireEye 2013; Ponemon Institute 2014a; Ponemon Institute 2014c).

---

[1] **Please cite this paper as**: *George Grispos, William Bradley Glisson, Tim Storer (2015). Security Incident Response Criteria: A Practitioner's Perspective. The 21st Americas Conference on Information Systems (AMCIS 2015), Puerto Rico, USA.*

Increasing economic pressures and regulatory forces, along with the apprehensions identified in industry reports, prompted the investigation into security incident response procedures from a practitioner's perspective. The contribution of this paper is a detailed case study of the security incident response challenges that face an organization. The case study is comprised of a set of interviews conducted with a Global Fortune 500 organization's Security Incident Response Team (SIRT). The interview results were combined with relevant literature to identify criteria which can be used to evaluate an organization's incident response capability and, as a guide, to support future security incident response improvement initiatives.

The remainder of this paper is structured as follows. Section two discusses security incident response approaches. Section three presents the methodology used in this research. Section four examines the results of the interviews. Section five presents an analysis of the organization's interview results and identified criteria. Section six draws conclusions and presents future research.

## Literature Review

Organizations have been exploring numerous solutions for implementing security incident response. As a result, several incident response approaches and best practice guidelines have been published in both industry (British Standards Institution 2011; Cichonski et al. 2012; European Network and Information Security Agency 2010) and academia (Mandia et al. 2003; Mitropoulos et al. 2006). Many of these approaches are based on a linear plan-driven model. Within these models, preparation for incident handling leads to incident detection and containment. In turn, a security incident response team can eradicate and recover from the incident and provide feedback into the wider organizational security posture. However, researchers have noted that there is a lack of consensus as to the standardization of security incident response (Alberts et al. 2004; Mitropoulos et al. 2006). As a result, there is currently no single de-facto approach which can truly be classified as an industry standard for handling security incidents (Mitropoulos et al. 2006).

Hove, et al (2014) studied three large organizations in an attempt to investigate the plans and procedures for handling security incidents within the studied organizations. The results from the study showed that although the organizations have plans and procedures in place, based on industry best practices, many procedures were missing from the organizational structure. The authors identified two organizations in which security incident reporting procedures were not established, while the respondents in another organization indicated that staff deficiencies impeded efficient response to incidents (Hove et al. 2014).

Werlinger, et al. (2007) conducted an exploratory study to investigate the security incident activities of practitioners in various organizations. The objective of the study was to determine what skills, tools and strategies were required to manage and handle security incidents. The results showed that practitioners often used pattern recognition and hypothesis generation during the analysis of security incidents. In a separate study, Werlinger, et al. (2010) added that current security incident response tools do not appropriately support the highly collaborative nature of incident investigations and that incident handlers often need to develop their own tools to perform specific tasks.

Line, et al (2014) examined how distribution service operators within the power industry planned and prepared for security incidents. The findings showed that many of the surveyed organizations had little or no documentation regarding the investigation of security incidents. Furthermore, Line, et al (2014) reported that the majority of their studied organizations did not have a clear definition for a security incident. This is a finding that is shared by Tan, et al (2003), who have also reported that the organization in their case study did not have a clear definition for a security incident. Tan, et al also noted that their organization was unaware of the benefits associated with collecting forensic evidence (Tan et al. 2003).

Casey (2006) argued that digital forensic techniques are increasingly being used in corporate environments during the investigation of information security incidents. Pangalos, et al. (2010) add that systems administrators and security incident response teams can use digital artifacts to analyze abnormal and malicious activities on computer networks, information systems and applications. However, several researchers (Casey 2005; Grobler and Louwrens 2007; Pangalos et al. 2010) have argued that organizations may not be fully exploiting their digital forensic capabilities and are likely to be undermining the value of forensic evidence. While previous security incident research has identified and discussed numerous challenges with current security incident response approaches, minimal empirical research investigates the identification of security incident response criteria.

## Methodology

This research empirically investigates security incident response practices and procedures by conducting an exploratory case study in a Global Fortune 500 organization. This exploratory case study was completed between November and December, 2013. The organization was selected for the case study due to a pre-existing relationship with the authors and their willingness to participate in the research domain. The name of the organization is being withheld to ensure organizational anonymity. Therefore, the names of organizational documents and processes have also been altered and the results of the interviews are presented anonymously. Maintaining organizational anonymity helps attain sensitive information while creating an environment that is conducive for all parties to present, potentially, sensitive information.

This study applied semi-structured interviews as the primary method of data collection (Oates 2006). The interviews were used to acquire a qualitative in-depth understanding of security incident response challenges faced in the context of a large organization. The survey instrument (available in the Appendix) consisted of a combination of open-ended and closed questions which were derived from themes identified in the literature. To mitigate researcher bias in terms of reliability and viability, the survey instrument was validated by two professionals in the organization (Kitchenham and Pfleeger 2002). An information security manager and a senior security analyst validated the instrument by taking the interview and providing feedback. The feedback from these individuals ranged from simplifying open-ended questions to adding response options to closed questions. It should be noted that this validation was only conducted once due to time constraints. Interviews were conducted in a conference room within the organization. Each interview lasted, approximately, 45 minutes with the interviewee's responses being recorded by hand. All hand-written notes were digitally documented within, roughly, one hour of the completed interview. In total, fifteen individuals were surveyed. Data analysis was performed using an inductive analysis approach which allows for research findings to "emerge from the frequent, dominant, or significant themes" identified within the raw survey data (Thomas 2006).

In addition to the interviews, an analysis of relevant literature was conducted to identify security incident response concerns established in other organizations. The literature was collected from academic papers, whitepapers and industrial surveys. The literature analysis helps to establish the global security incident response perspective and support derived criteria.

## Interview Analysis

The initial survey questions establish the interviewee's current role within the organization and quantified years of experience in information technology. The answers from these questions revealed that the interviewees have on average of thirteen and a half years' experience within an information technology role. More specifically, the individuals identified themselves as information security managers, senior security analysts or security analysts who assume various duties within the organization.

The following sub-sections define the organization's security incident response process and present an analysis of the interview results. These results examine the respondent's perception of security incident response, data gathering within the process, challenges to security incident response as well as incident learning and dissemination.

*Security Incident Response Process*

In order to comprehend how a Security Incident Response Team (SIRT) functions, initial questions examined the organization's security incident response process. The incident response process findings that are of particular interest to this research are as follows:

- The organization uses a customized linear document-centric security incident response approach which consists of four phases: incident detection and reporting; recording, classification and assignment; investigation and resolution; and incident closure.
- The surveyed individuals indicated that the documented process is not always followed. When queried as to the reasons for deviating from the process, answers included time constraints, a lack of staff to run the entire process, a lack of support for handling specific incidents and a rigid, heavy document-centric approach to processing an incident.
- The general indication from the participants is that the documented process provides structure to the SIRT. The documentation offers insight into how the SIRT can resolve incidents and provides clarity on the escalation path to other business units.
- The results revealed that the current documented process is not dynamic when considering quick resolution, and lacks detail regarding the level of information that is required to be recorded for individual incidents.

*Perception of Security Incident Response*

Interviewees were queried as to what the term 'security incident' meant to them. A wide variety of answers were received, which included "a breach of security policy", "a degradation of security controls", "data loss", "financial losses" and "a threat to service availability". The variety of answers received from this query indicates that there is lack of cohesion on how a security incident is viewed within the organization. This can also indicate that the organization does not have a unified definition for the term 'security incident'.

Although all the participants have been involved at some stage in the management of an incident, when asked specifically if the organization has a SIRT, thirteen out of the fifteen respondents indicated that there is no dedicated team. Actually, the organization has what can be described as an ad-hoc SIRT, where individuals are brought together to investigate particular incidents.

Interviewees were then asked about which phases of the security incidents response process they have been actively involved in within the organization. The results of that inquiry are summarized in Table 1 - Incident Response Involvement. The phases provided to the respondents were derived from the literature (Cichonski et al. 2012; Mitropoulos et al. 2006). The table indicates that the majority of participants reported that they were involved in the identification, eradication, investigation, and recovery phases of incident response.

When asked if there is a documented security incident response process, twelve out of the fifteen individuals indicated that there is a documented process for handling incidents. However, when asked to recall the process, only five out of the twelve positive respondents could actually recall the process itself. The organization does have a documented security incident response process; however immediate mental recollection of the details in the documented process was limited to a few individuals.

| Phase Name | Number of Participants |
|---|---|
| Preparation | 5 |
| Identification | 9 |
| Containment | 7 |
| Eradication | 9 |
| Investigation | 13 |
| Recovery | 10 |
| Follow-up | 4 |

**Table 1 - Incident Response Involvement**

### *Data Gathering within the Process*

When asked if the organization collected information about an incident during an investigation, the majority of the answers returned were positive. Fourteen out of the fifteen individuals indicated that information about security incidents is collected and stored during the response lifecycle. One individual indicated that he/she 'does not know' if the practice took place. The respondents indicated that incident information collection was usually assigned and performed by the primary incident handler. This individual is given the responsibility to collect and record this information. The information typically recorded for security incidents includes incident meeting notes, actions to be taken for remediation, copies of logs, as well as email communication between the SIRT and management. Documentation is usually tailored to specific incidents and there does not appear to be a uniform approach to capturing specific information. It appears that the general practice is to capture information which is required to eradicate and recover from the incident which may not necessarily facilitate incident learning at a later stage.

The respondents indicated that the organization uses two databases to store information related to security incidents. One database is used to record intrusion detection system incidents, while the other database is used to record all other security incidents. However, respondents indicated that access to these databases was limited to only a subset of individuals within the information security unit. Participant responses indicate that there are opportunities for process improvements. The enhancements proposed by these individuals included, making the database more searchable, providing additional guidance with regard to what information to record, as well as implementing a 'lightweight' incident record.

The interviews confirmed that the above information was not the only information being collected about an incident. The SIRT also gathers forensic data from various sources. This can include logs, emails, hard disk drive images, and physical memory dumps. This type of information can be used as evidence in legal proceedings, if the need arises. Individuals were asked if there was a process to collect this information, only five individuals indicated that there was such a process. One individual indicated that no such process exists and nine 'did not know' if such a process existed within the organization. Documented processes do exist within the organization, which describe methods for acquiring data from hard drives, the storage of information in a secure location, as well as an E-Discovery process. However, the results indicate that there is limited diffusion of information relating to process knowledge and tool access.

If an organization is going to collect forensic data for use in possible legal action, then a chain of custody process should be considered (Casey 2011). The five respondents who provided positive answers to the existence of a forensic data collection process were then queried if the chain of custody practice was performed. Two individuals specified that this practice exists and was done for all incidents. However, two other individuals noted that a chain of custody process does exist but was not performed all the time. There was one 'do not know' answer. The organization does have a defined chain of custody process which specifies how an incident handler acquires and stores forensic data to show a continuous chain of custody. The two individuals, who noted that the chain of custody was not performed consistently, suggested that further guidance was required as to when this should be undertaken.

### *Challenges to Conducting Security Incident Response within the Organization*

One of the goals of the survey was to establish the challenges facing a SIRT within an organization. The interviews highlighted three main challenges for a SIRT. First, the process for employee reporting and escalation of security incidents needs to be clearly defined, established, implemented and, periodically, refined as processes and technology evolve. Second, the establishment of comprehensive data access for incident response team members to perform investigations. Third, resolving conflict resolution between the SIRT and the business units responsible for the availability of customer-facing assets.

The respondents were asked how security incidents are reported within the organization. Three individuals stated that there is a documented process for reporting security incidents, six indicated that no such process exists within the organization, and six provided a 'do not know' answer. From the respondent's answers, it appears that the majority of security incidents are reported informally either verbally or via email, usually to an associate within the information security unit. However, the respondents generally agreed that the security incident reporting process could be improved. The respondent's main concern with the current approach is that when certain members of the information security team were unavailable, incidents can take longer to reach the SIRT. Other potential improvements proposed include an incident reporting hotline and a dedicated email address for reporting incidents directly to the SIRT.

Ten out of the fifteen respondents indicated that the SIRT often has difficulties conducting investigations due to a lack of access to security data. There were a variety of answers describing the obstacles preventing investigation including limited physical access to security data, short data retention times, logs not containing enough detailed information and limited support from third-parties involved in incidents.

The respondents also indicated that conflicts arise between the SIRT and various other stakeholders within the organization. From time-to-time, security incidents which affect the availability of customer-facing assets can lead to a disagreement over returning the asset back to the production environment and performing a more complete security investigation. This conflict originates from the fact that the organization relies on the continuous availability of customer-facing and back-office applications. The respondents also stated that conflicts can occur when there is a lack of physical access to systems and applications to extract security data for security investigations. However, the majority of respondents indicated that both conflicts are resolved at a management level.

### *Incident Learning and Dissemination*

Researchers (Ahmad et al. 2012; Tan et al. 2003) have suggested that many organizations are not maximizing their post-incident learning potential and tend to focus on only improving technical processes in an attempt to prevent reoccurrence. When the interviewees were asked if the organization performs any 'post-incident' activities, ten out of the fifteen respondents indicated that depending on the type of incident, several activities can take place. These activities are in line with the findings of previous researchers, which include implementing security controls to prevent reoccurrence, producing reports for management and education awareness through policy reiteration. There were four 'do not know' answers and one individual stated that no 'post-incident' activities took place within the organization.

In order to investigate if any further learning was taking place during the organization's incident process, the respondents were queried to determine if a root cause analysis was performed post-incident. Seven out of the ten respondents indicated that this was the case within the organization. The three remaining respondents indicated that they have not been involved in incidents where a root cause analysis was required. It was interesting to note that one respondent indicated that a root cause analysis should be done for each incident, but they have been involved in incidents where this activity was not performed.

All ten respondents who indicated that the organization performs 'post-incident' activities suggested that there is the potential to enhance and extend these post-incident activities in order to focus on improving the effectiveness of internal policies, procedures, controls and training. A number of recurring themes were mentioned as potential enhancements. These included a deeper analysis of security incidents, improving methods to assist in the development of lessons learned, implementing security controls focused more on preventing incident reoccurrence, as well as, increasing the dissemination of lessons learned.

Ahmad, et al (2012) argued that the work of a SIRT is usually completed by the issuance of a report, detailing the investigation findings and any lessons learned along with the distribution of information internally or externally. When the interviewees were asked if any post-incident information was distributed or disseminated to any groups or departments within the organization, six out of the fifteen respondents said 'yes'. Two individuals said that post-incident information was not distributed or disseminated within the organization and seven individuals 'did not know' if this took place. The six respondents indicated that there are two post-incident information dissemination methods. First, an electronic announcement on the organization's Intranet and second through a monthly statistics collected for management. However, when queried if a process exists for distributing this information, the respondents were unanimous that there is no formal process for distributing this information within the organization. The respondents did indicate that the decision to distribute incident information is held with the information security manager.

The respondents were also asked if any post-incident information was distributed or disseminated outside of the organization, for example to regulatory bodies. Nine out of the fifteen respondents indicated that post-incident information was distributed or disseminated outside of the organization. One individual suggested that this did not take place and five individuals noted they 'do not know' if the practice was taking place. When asked if there was a process in place to govern this practice, all nine respondents indicated that they 'did not know' if such a process existed. One respondent did note that if such actions were required, the Regulatory Compliance Unit would disclose the incident to the relevant regulators.

## Security Incident Response Criteria (SIRC)

Industry surveys and the relevant literature have established the global problem with security incident response, while the Fortune 500 organization interview results were used to identify local security incident response problems. A set of criteria has been derived from relevant literature and the survey results to improve security incident response processes. SIRC identifies six essential criteria that a successful security incident response process needs to address:

- Dynamic stakeholder involvement
- Multidisciplinary security incident response team
- Short investigation lifecycle times
- Incident learning throughout the incident lifecycle
- Access to security data
- Protecting digital evidence

### *Dynamic Stakeholder Involvement*

Stakeholder support for investigating and learning from security incidents is critical. Without the support of stakeholders, a Security Incident Response Team (SIRT) cannot perform a complete investigation. Stakeholder support for security incident response needs to be both proactive and reactive. Stakeholders need to be proactive by giving them appropriate tools, access to data and the support to be successful in their investigations. Similarly, stakeholders need to be reactive by encouraging incident learning during investigations while ensuring assets are not offline for an unnecessary extended period of time. Reactive support includes encouraging process communication among SIRT members. This is evident within the organizational interviews, where the results indicate that there is limited diffusion of information relating to process knowledge and tool access.

Stakeholders can also encourage security communication among employees within an organization. For example, employees should be encouraged to report even the smallest of security incidents to the incident response team without fear of repercussion. Increased trust by employees to report security incidents in a timely manner, potentially, translates into a better understanding of how incident learning can improve the wider organizational security posture.

### *Multidisciplinary Response Team*

A recent industry survey has suggested that security incident response is becoming an increasingly multidisciplinary affair, usually involving a number of individuals with a variety of skills (Ponemon Institute 2014a). In addition, Werlinger, et al (2007) observed that SIRTs often collaborate with various individuals and multidisciplinary groups within their organizations.

The responses from the interviews support previous observations and findings. Respondents in the organizational interviews indicated that other organizational units not associated with the information security unit are routinely involved in the investigation of security incidents. These units assist in the decision-making process to investigate specific assets. Also, they are required to provide access to security data and support SIRT for specific incidents, such as E-discovery requests.

The conflicts identified in the interviews between the goals of the SIRT and the desires of business units to return customer-facing assets online as quickly as possible, supports the idea for multidisciplinary response teams. The idea is to encourage increased communication among all parties with a vested interest in resolving an incident. A multidisciplinary security incident response team could require the integration of technical security professionals and information technology specialists, with relevant asset stakeholders, while closely working with an organization's legal department. An important aspect of this multidisciplinary incident response team is that members must understand their roles and responsibilities within the team.

### *Short Investigation Lifecycles Times*

Time is a critical factor in security incident investigations (Ahmad et al. 2012; Mitropoulos et al. 2006). The ability to swiftly respond and investigate why a security incident has occurred can reduce system downtime, subsequent financial losses, as well lowering the cost of returning the organization to its normal security posture (Killcrece et al. 2003).

Within the organizational interviews, participants indicated that a time-related conflict occurs between the security incident response team and business units responsible for the continuous availability of information assets. The participants indicated that the conflict resolves around desire to investigate the security elements of an incident and the necessity to restore the affected asset to satisfy business requirements. Johnson (2013a) has noted that this conflict can also occur in security incident investigations involving safety-critical applications. Therefore, a security incident response process needs to address the time pressures associated with returning information assets to the production environment and quicker incident resolution through shorter investigation lifecycles.

### *Learning Lifecycle*

Information security incidents are unwanted occurrences, yet at the same time they present an opportunity to learn about the risks and vulnerabilities which can exist in both technical and socio-technical systems (Line et al. 2009). Any lessons learned from security incidents and security incident handling processes can then be used by an organization to improve its wider information security posture. However, researchers have argued that organizations do not pay enough attention to incident learning (Ahmad et al. 2012; Shedden et al. 2011). These researchers go on to claim that organizations are more concerned with eradication and recovery (Ahmad et al. 2012; Shedden et al. 2011).

The integration of security incident learning throughout security incident response processes, potentially, address this issue. The integration of incident recalls, identifying event/incident sequences, performing root cause and barrier analysis, along with information dissemination methods, can all encourage incident learning throughout the incident lifecycle. Incident recalls can occur at the start of the incident lifecycle, allowing an incident handler to identify process improvements based on previous experience with similar incidents. Similarly, identifying event/incident sequences at the start of the lifecycle, allows an incident response team to increase the amount of data potentially being captured, therefore enhancing the technical excellence of incident learning (Grispos et al. 2014). Incident learning techniques such as a root cause and barrier analysis, during an incident investigation can assist in identifying the technical and socio-technical issues that have contributed to the incident (Johnson 2013a). These forms of analysis could be used to establish why a security incident has occurred, as well as identifying the barriers that should have been in place to aid its prevention (Johnson 2013a). The end-products of any incident learning can then be distributed at the end of any investigation.

### *Access to Security Data*

Security incident investigations can provide information that can be instrumental in avoiding a recurrence of the incident. However, a SIRT cannot establish the technical and underlying causes of a security incident without access to enriched security data. Ten out of the fifteen respondents in the organizational interviews indicated that limited data access to affected assets, data ownership issues, data life span and/or limited data capture configurations have hindered enriched incident learning.

The widespread use of digital technology in organizations has led to an abundance of data sources for SIRTs (Cichonski et al. 2012). Access to detailed data from these sources can increase the potential for a SIRT to establish incident root causes, as well as identifying information that could be used to prevent future occurrences (Lindberg et al. 2010). A security incident response process should, therefore, include provisions for an incident response team to access richer information to perform a more complete investigation.

### *Protecting Digital Evidence*

Industrial surveys have suggested that organizations are increasingly impacted by incidents which contain a criminal element (FireEye 2013; PricewaterhouseCoopers 2014). As a result, digital forensics techniques are increasingly being used in corporate environments during the investigation of security incidents (Casey 2006). However, literature suggests that many organizations only involve forensic principles when it may be too late into the investigation (Mitropoulos et al. 2006). The survey results indicate that although the process exists, a chain of custody was not performed at all times during evidence collection. The collection of evidence should be performed in a 'forensically sound' manner to ensure that complete and accurate copies of data are obtained, and the authenticity of the evidence is documented for future reference. Proper documentation and a chain of custody can provide a strong indication that digital evidence has been handled properly and has not been altered or contaminated.

## Conclusions

Information security incidents are increasingly impacting organizations. As a result, security incident response measures are being implemented to detect, respond to, and recover from these incidents. However, a real-world investigation of security incident response reveals that it is a complicated issue in an increasingly challenging environment. These challenges can include balancing the security investigation of assets with their swift return to the production environment and prompting security incident learning. The conclusion derived from the organizational interviews and previous academic research is that organizations could benefit from an alternative approach to handling and managing security incidents. The interview data coupled with the literature review supports the identification of the Security Incident Response Criteria (SIRC).

Future work will consider an examination of the SIRC against existing security incident response methodologies to determine security incident response applicability. Future work will also examine the development of an alternative security incident response methodology based on the SIRC. A possible solution for this methodology could be the integration of disciplined agile principles and practices into the security incident response process. This will involve investigating the incorporation of the SIRC with agile principles to provide the foundation for the development of an Agile Incident Response methodology. Ultimately, the high-level goal is the deployment and continued refinement of an agile security incident response methodology.

## Acknowledgements

This work was supported by the A.G. Leventis Foundation. Any opinions, findings, conclusions or recommendations expressed in this paper are those of the authors and do not reflect the views of the A.G. Leventis Foundation.

**Appendix**

*The interview instrument which guided this research is presented below.*

1. What is your current job title/role?

2. How many years have you worked in IT?

3. Briefly describe the key areas of your job function/role?

4. From your perspective, what is meant by the term 'security incident'?
5. Does the organization have a security incident response team? YES/NO/Don't Know (DNK)
    a. If YES, what are the overall goals of the security incident response team?
    b. If NO, are there plans to develop one in the future? YES/NO/DNK
        i. If YES, what is the projected time frame?
        ii. If NO, is there a reason for not developing a security incident response team?
6. Does the organization have a documented security incident response process? YES/NO/DNK
    *a.* If YES, can you briefly describe this documented process and the activities involved in the process? *In the event participant cannot recall the exact process, he/she was asked to recall to the best of their ability any phases of the process to stimulate discussion.*
    b. If YES, in your opinion, does this documented process ensure that the goals of the security incident response team are met?
    c. If YES, in your opinion what are the good points in this documented process?
    d. If YES, in your opinion what are the bad points in this documented process?
    e. If YES, have you ever found it necessary to deviate from this documented incident response process? YES/NO/DNK
        i. If YES, why?
    f. If NO, are there plans to develop one in the future? YES/NO/DNK
        i. If YES, what is your projected time frame?
        ii. If NO, is there a reason for not developing such a documented process in the organization?

7. Is there an individual in the organization who is accountable for the documented security incident response process not being followed? YES/NO/DNK
    a. If YES, what is his/her title? Is this person accountable for legislative compliance as well? YES/NO/DNK
        i. If NO, who in the organization is accountable for legislative compliance?
    b. If NO, why not?

8. In your opinion, what are an individual's responsibilities for each the following categories in the security incident response lifecycle?
    a. Preparation
    b. Identification
    c. Containment
    d. Eradication
    e. Investigation
    f. Recovery
    g. Follow-up (Post-incident)

9. Are you involved in any of the above categories of the security incident response lifecycle? YES/NO
    a. If YES, which categories?

10. In your opinion, do you think that the documented security incident response process could be improved? YES/NO/DNK
    a. If YES, how and why?
    b. If NO, why not?

11. Does the organization have a documented procedure outlining how security incidents are reported within the organization? YES/NO/DNK
    a. If YES, what is this procedure?
    b. If YES, in your opinion, is this security incident reporting procedure effective? YES/NO/SOMETIMES/DNK
        i. If YES, why?
        ii. If NO, why not?
        iii. If SOMETIMES, when is it effective?
        iv. If SOMETIMES, when is it not effective?
    c. If NO, how are security incidents reported?
    d. Are employees within the organization educated on how to report security incidents? YES/NO/DNK

12. Does the organization record and store information related to security incidents? YES/NO/DNK
    a. If YES, can you briefly describe this practice and what information is recorded for security incidents?
    b. If YES, in your opinion, is the practice used for recording information related to security incidents effective? YES/NO/SOMETIMES/DNK
        i. If NO, why not?
        ii. If SOMETIMES, when is it effective?
        iii. If SOMETIMES, when is it not effective?
    c. If YES, where are these incidents recorded?
    d. If YES, who has access to these incident 'reports' (groups and departments)?
    e. If YES, is this procedure used to document and recorded all types of security incidents? YES/NO/DNK
        i. If NO, which incidents types are not recorded using this procedure?
        ii. If NO, where are these incidents recorded?
    f. If NO, why not?

13. Is there a documented process for collecting 'forensic data' related to security incidents such as (forensic images, logs etc.)? YES/NO/DNK
    a. If YES, can you briefly describe this process?
    b. If YES, does this process include a 'chain of custody' document or is an audit trail maintained? YES/NO/DNK
        i. If YES, can you briefly discuss this process?
        ii. If NO, why not?
    c. If NO, are there plans to develop one in the future? YES/NO/DNK
        i. If YES, what is your projected time frame?

ii. If NO, is there a reason for not developing such a documented process in the organization?

14. In your experience, do situations occur where an incident handler does not have access to all the information he/she requires (forensic images, logs etc.), when investigating security incidents? YES/NO/DNK
    a. If YES, what types of situations?
    b. If YES, what information was required?
    c. If NO, do you for see this as a problem in the future? YES/NO/DNK
        i. If YES, in what way?
        ii. If NO, why do you think this will not be a problem in the future?

15. In your experience, do conflicts arise between different stakeholders within the organization when investigating a security incident? YES/NO/DNK
    a. If YES, what are these conflicts?
    b. If YES, how are these conflicts resolved?

16. In your experience, what constitutes the closure of an information security incident within the organization?
    a. Can you recall an instance(s) where an incident has not been closed? YES/NO/DNK
        i. If YES, why was the incident not closed?
    b. Can you recall an instance(s) where an incident has been reopened? YES/NO/DNK
        i. If YES, why was the incident reopened?

*17.* From your experience, does the organization perform any 'post-incident' activities after a security incident has been closed? YES/NO/DNK *In the event participant answers NO or DNK, they were asked questions 17a-d seeking their opinions on what these activities should involve, the idea is to promote discussion around the topic of post-incident activities.*
    a. If YES, what activities take place 'post-incident'?
    b. If YES, in your opinion do you think these activities are sufficient? YES/NO/DNK
        i. If NO, what 'post-incident' activities do you think should occur?
    c. If YES, do these activities include a root cause analysis? YES/NO/DNK
        i. If NO, why not?
    d. If YES, do these activities include a risk analysis based on the incident? YES/NO/DNK
        i. If NO, why not?
    e. If NO, why not?

18. Does the organization distribute any 'post-incident' information to any group(s) or department(s) within the organization? YES/NO/DNK
    a. If YES, is there a formal process for disseminating this 'post-incident' knowledge? YES/NO/DNK
        i. If YES, can you briefly describe this process?
        ii. If NO, why not?

b. If YES, is there an individual in the organization who is responsible for disseminating 'post-incident' knowledge within the organization?
       i. If YES, what is his/her title?
       ii. If NO, why not?
   c. If YES, to which groups or departments?
   d. If YES, what type of information is circulated?
   e. If NO, why not?

19. Does the organization distribute any 'post-incident' information or incident notifications outside of the organization, for example to regulatory bodies? YES/NO/DNK
    a. If YES, is there a formal process for disseminating this 'post-incident' knowledge? YES/NO/DNK
        i. If YES, can you briefly describe this process?
        ii. If NO, why not?
    b. If YES, is there an individual in the organization who is responsible for disseminating 'post-incident' knowledge outside the organization?
        i. If YES, what is his/her title?
        ii. If NO, why not?
    c. If YES, where is this information circulated?
    d. If YES, what type of information is circulated?
    e. If NO, why not?

20. Are you aware of any individual in the organization who is responsible for analysing previous security incidents to identity trends, anomalies, and patterns? YES/NO/DNK
    a. IF YES, what is his/her title?
    b. If NO, how are trends, anomalies, and patterns identified from security incidents?

21. Does the organization consult or contract any individuals/contractors/businesses to help detect, investigate, eradicate or recover from a security incident? (i.e. forensic analysis work) YES/NO/DNK
    a. If YES, can you briefly describe an instance where an individual/contractor/ business has been included in the handling of an incident?
    b. If YES, do individuals/contractors/businesses follow the same security incident response process as employees? YES/NO/DNK
    c. If NO, do you think this will occur in the future?

22. Does the organization have a documented secure development process (SDLC) or like-policies around the secure development of systems and applications? YES/NO/DNK
    *In the event participant answers DNK, they were asked question 22d to promote discussion around the idea or integrating lessons learned into the application development environment.*
    a. If YES, can you briefly describe the organization's secure development process?
    b. If YES, in your opinion what are the good points in this documented process?
    c. If YES, in your opinion what are the bad points in this documented process?

d. If YES, in your opinion, do you think that 'lessons learned' from previous security incidents should play a role in the secure development process? YES/NO/DNK
      i. If YES, why so?
      ii. If NO, why not?
   e. If NO, why not?

23. Does the organization have an information security education program? YES/NO/DNK
    a. If YES, can you briefly highlight the purpose of the program?
    b. If YES, do you know if any information from previous security incidents is included in the security education program? YES/NO/DNK
       i. If YES, can you briefly highlight what this information contains?
       ii. If NO, are there plans to include such information in the future? YES/NO/DNK
    c. If NO, are there plans to develop one in the future? YES/NO/DNK
       i. If YES, what is your projected time frame?
       ii. If NO, is there a reason for not developing an information security education program in the organization?

24. Were any of the questions vague or difficult to follow?

25. Do you have any additional information you would like to add? Is there anyone else you suggest I talk to regarding any information in this survey?